\documentclass[12pt]{article}

\newcommand{\beg}{\begin{equation}}
\newcommand{\ene}{\end{equation}}

\begin{document}
\title{
\textsc{\bf Einstein, Podolsky and Rosen}\\
\textsc{\bf versus}\\
\textsc{\bf Bohm  and Bell} }

\author{
Andrei Khrennikov and Igor Volovich
\\ $~~~$\\
\textsf{International Center in Mathematical Modelling in }\\
\textsf{Physics, Engineering and Cognitive Sciences}\\
\textsf{University of Vaxjo, S-35195, Sweden}\\
 $~~~$\\
\textsf{Steklov Mathematical Institute}\\
\textsf{Russian Academy of Sciences}\\
\textsf{Gubkin St. 8, 117966, GSP-1, Moscow, Russia}\\
\emph{e-mail: volovich@mi.ras.ru} }
\date {~}
\maketitle
\begin{abstract}
There is an opinion that the Bohm reformulation of the EPR paradox
in terms of  spin variables is equivalent to the original one.
In this note we show that such an opinion  is not justified. We
apply to the original EPR problem the method which was used by
Bell for the Bohm reformulation. He has shown that  correlation
function of two spins  cannot be represented by classical
correlation of separated bounded random processes. This Bell`s
theorem  has been interpreted as incompatibility of local realism
with quantum mechanics. We  show that, in contrast to Bell`s
theorem for spin or polarization correlation functions, the
correlation of positions (or momenta) of two particles, depending
on the rotation angles, in the original EPR model always
admits a representation in the form of classical correlation of
separated random processes. In this sense there exists a local
realistic representation for the original EPR model but there
is no such a representation for the Bohm spin reformulation of the
EPR paradox. It shows also that the phenomena of quantum
nonlocality is based not only on the properties of entangled
states but also on the using of particular bounded observables.
\end{abstract}
\newpage



In 1935 Einstein, Podolsky and Rosen (EPR) advanced an argument
about incompleteness of quantum mechanics \cite{EPR}. They
proposed a gedanken experiment involving a system of two particles
spatially separated but correlated in position and momentum and
argued that two non-commuting variables (position and momentum of
a particle) can have simultaneous physical reality. They concluded
that the description of physical reality given by quantum mechanics, 
which does not permit such a simultaneous reality,
is incomplete.

Though the EPR work dealt with continuous variables most of the
further activity have concentrated almost exclusively on systems
of discrete spin variables following to the Bohm \cite{Boh} and
Bell \cite{Bel1} works.


Bell's theorem~\cite{Bel1} states that there are quantum spin
correlation functions that can not be represented as classical
correlation functions of separated  random variables. It has been
interpreted as incompatibility of the requirement of locality with
the statistical predictions of quantum mechanics~\cite{Bel1}. For a
recent discussion of Bell's theorem see, for example ~\cite{CS} -
~\cite{Vol2} and references therein.
 It is now widely accepted, as a result of
Bell's theorem and related experiments, that "Einstein`s local
realism" must be rejected. For a discussion of the role of
locality in the three dimensional space see, however, \cite{Vol1,Vol2}.

The original EPR system involving continuous variables has been
considered by Bell in \cite{Bel2}. He has mentioned that if one
admits "measurement" of arbitrary "observables" on arbitrary
states than it is easy to mimic his work on spin variables (just
take a two-dimensional subspace and define an analogue of spin
operators). The problem which he was discussing in \cite{Bel2} is
narrower problem, restricted to measurement of positions only, on
two non-interacting spinless particles in free space. Bell used
the Wigner distribution approach to quantum mechanics.The original
EPR state has a nonnegative Wigner distribution. Bell argues that
it gives a local, classical model of hidden variables and
therefore the EPR state should not violate local realism. He then
considers a state with nonpositive Wigner distribution and
demonstrates that this state violates local realism.

Bell`s proof of violation of local realism in phase space has been
criticized in \cite{Joh} because of the use of an unnormalizable
Wigner distribution. Then in \cite{BW} it was demonstrated that
the Wigner function of the EPR state, though positive definite,
provides an evidence of the nonlocal character of this state
if one measures a displaced parity operator.

In this note we apply to the original EPR problem the method which
was used by Bell in his well known paper \cite{Bel1}. He has shown
that the correlation function of two spins  cannot be represented
by classical correlations of separated bounded random variables.
This Bell`s theorem  has been interpreted as incompatibility of
local realism with quantum mechanics. We shall show that, in
contrast to  Bell`s theorem for spin correlation functions, the
correlation function of positions (or momenta) of two particles
always admits a representation in the form of classical correlation
of separated random variables. This result looks rather surprising
since one thinks that the Bohm-Bell reformulation of the
EPR paradox is equivalent to the original one.

In the Bohm formulation one considers a pair of spin one-half
particles formed in the singlet spin state and moving freely
towards two detectors. If one neglects the space part of the wave
function then one has the Hilbert space $C^2\otimes
C^2$ and  the quantum mechanical correlation of two spins
in the singlet state $\psi_{spin}\in C^2\otimes
C^2$ is
\begin{equation}
 D_{spin}(a,b)=\left<\psi_{spin}|\sigma (a) \otimes\sigma
(b)|\psi_{spin}\right>=-a\cdot b \label{eq:eqn1}
\end{equation}
Here $a=(a_1,a_2,a_3)$ and $b=(b_1,b_2,b_3)$ are two unit vectors
in three-dimensional space $R^3$,
$\sigma=(\sigma_1,\sigma_2,\sigma_3)$ are the Pauli matrices,
 $
 \sigma (a) =\sum_{i=1}^{3}\sigma_i a_i
 $
 and
$ \psi_{spin}= (|01>-|10>/\sqrt 2. $

{\it Bell's theorem} states that the function $ D_{spin}(a,b)$
Eq.~(\ref{eq:eqn1}) can not be represented in the form
\begin{equation}
\label{eq:eqn2} \int \xi_1 (a,\lambda) \xi_2 (b,\lambda)
d\rho(\lambda)
\end{equation}
i.e.
\begin{equation}
\label{eq:Ab} D_{spin}(a,b)\neq \int \xi_1 (a,\lambda) \xi_2
(b,\lambda) d\rho(\lambda)
\end{equation}
Here $ \xi_1 (a,\lambda)$ and $  \xi_2(b,\lambda)$ are random  fields
on the sphere, which satisfy the bound
\begin{equation}
\label{eq:Abb}
 |\xi_n (a,\lambda)|\leq 1,~ ~ n=1,2
\end{equation}
 and $d\rho(\lambda)$ is a positive probability
measure,  $ \int d\rho(\lambda)=1$. The parameters $\lambda$ are
interpreted as hidden variables in a realist theory.

 The proof of
the theorem is based on Bell`s (or CHSH) inequalities. Let us
stress that the main point in the proof is actually not the
discreteness of  classical or quantum spin variables but the
bound  (\ref{eq:Abb}) for classical random fields.

If we relax the bound then one can exibit a local realistic model
which reproduces quantum correlation of two spins. Indeed 
 let us take as a
probability space $\Lambda$ just 3 points: $\Lambda =\{1,2,3\}$
and the expectation
$$
Ef=\frac{1}{3}\sum_{\lambda =1}^3f(\lambda)
$$
Let the random fields be
\begin{equation}
\label{eq:Bcl}
\xi _1(a,\lambda)=-\xi_2 (a,\lambda)=\sqrt 3 a_\lambda,~~
\lambda=1,2,3
\end{equation}

Then one has the relation:

\begin{equation}
\label{eq:Bcl2}
-(a,b)=E\xi_1(a)\xi_2(b)
\end{equation}

The Bell`s theorem (\ref{eq:Ab}) does not valid in this case because
we do not have the bound (\ref{eq:Abb}). Instead we have
\begin{equation}
\label{eq:Bcl3}
|\xi_n (a,\lambda)|\leq \sqrt 3
\end{equation}

Now let us apply a Bell style  approach through correlation functions
 to the original EPR case.
The Hilbert space of two one-dimensional particles is
$L^2(R)\otimes L^2(R)$ and canonical coordinates and momenta are
$q_1,q_2,p_1,p_2$ which obey the commutation relations
\begin{equation}
\label{eq:EPR1}
[q_m,p_n]=i\delta_{mn},~~[q_m,q_n]=0,~~[p_m,p_n]=0,~~m,n=1,2
\end{equation}

The EPR paradox can be described as follows. There is such a state
of two particles that  by measuring $p_1$ or $q_1$ of the first
particle, we can predict with certainty and without interacting
with the second particle, either the value of $p_2$ or the value
of $q_2$ of the second particle. In the first case $p_2$ is an
element of physical reality, in the second $q_2$ is. Then, these
realities must exist in the second particle before any measurement
on the first particle since it is assumed that the particle are
separated by a space-like interval. However the realities can not
be described by quantum mechanics because they are incompatible --
coordinate and momenta do not commute. So that EPR conclude that
quantum mechanics is not complete. Note that the EPR state
actually is not a normalized state since it is represented by the
delta-function, $\psi=\delta(x_1-x_2-a).$

An important point in the EPR consideration is that one can choose
what we measure -- either the value of $p_1$ or the value of
$q_1$. For a mathematical formulation of a free choice we
introduce canonical transformations of our variables:
\begin{equation}
\label{eq:EPR2} q_n(\alpha)=q_n\cos \alpha - p_n\sin \alpha,~~
p_n(\alpha)=q_n\sin\alpha + p_n\cos\alpha;~~n=1,2
\end{equation}
Then one gets
\begin{equation}
\label{eq:EPR3} [q_m(\alpha),p_n(\alpha)]=i\delta_{mn}; ~m,n=1,2.
\end{equation}
In particular one has $q_n(0)=q_n,~~q_n(3\pi/2)=p_n,~~n=1,2.$

Now let us consider the correlation function
\begin{equation}
\label{eq:EPR4} D(\alpha_1,\alpha_2)=\left<\psi|q_1 (\alpha_1)
\otimes q_2(\alpha_2)|\psi\right>
 \end{equation}
The correlation function $D(\alpha_1,\alpha_2)$ (\ref{eq:EPR4}) is
an analogue of the Bell correlation function  $D_{spin}(a,b)$
(\ref{eq:eqn1}). Bell in \cite{Bel2} has suggested to consider the
correlation function of just the free evolutions of the particles
at different times (see  below).

 We are interested in the
question whether the quantum mechanical correlation function
(\ref{eq:EPR4}) can be represented in the form
\begin{equation}
\label{eq:EPR5} \left<\psi|q_1 (\alpha_1) \otimes
q_2(\alpha_2)|\psi\right>=\int \xi_1 (\alpha_1,\lambda)
\xi_2(\alpha_2,\lambda) d\rho(\lambda)
\end{equation}
Here $ \xi_n (\alpha_n,\lambda), n=1,2$  are two real functions (random
processes), possibly unbounded, and $d\rho(\lambda)$ is a positive
probability measure, $ \int d\rho(\lambda)=1$. The parameters
$\lambda$ are interpreted as hidden variables in a realist theory.

Let us prove that there are required functions $ \xi_n
(\alpha_n,\lambda)$ for an arbitrary state $\psi$. We rewrite the
correlation function $D(\alpha_1,\alpha_2)$ (\ref{eq:EPR4}) in the
form
\begin{equation}
\label{eq:EPR6} \left<\psi|q_1 (\alpha_1) \otimes
q_2(\alpha_2)|\psi\right>=
<q_1q_2>\cos\alpha_1\cos\alpha_2-<p_1q_2>\sin\alpha_1\cos\alpha_2
\end{equation}
$$
-<q_1p_2>\cos\alpha_1\sin\alpha_2
+<p_1p_2>\sin\alpha_1\sin\alpha_2
$$
Here we use the notations as
$$<q_1q_2>=\left<\psi|q_1
q_2|\psi\right>$$
Now let us set
$$
\xi_1(\alpha_1,\lambda)=f_1(\lambda)\cos\alpha_1-g_1(\lambda)\sin\alpha_1,
$$
$$
\xi_2(\alpha_2,\lambda)=f_2(\lambda)\cos\alpha_2-g_2(\lambda)\sin\alpha_2
$$
Here real functions $f_n(\lambda),g_n(\lambda),~n=1,2$ are  such that
\begin{equation}
\label{eq:EPR7}
Ef_1f_2=<q_1q_2>,~Eg_1f_2=<p_1q_2>,~Ef_1g_2=<q_1p_2>,~Eg_1g_2=
<p_1p_2>
\end{equation}
We use for the expectation the notations as
$Ef_1f_2=\int f_1(\lambda) f_2,(\lambda) d\rho(\lambda)$. To solve
the system of equations (\ref{eq:EPR7}) we take
$$
f_n(\lambda)=\sum_{\mu =1}^2F_{n\mu}\eta_{\mu}(\lambda),~
g_n(\lambda)=\sum_{\mu =1}^2G_{n\mu}\eta_{\mu}(\lambda)
$$
where $F_{n\mu},G_{n\mu}$ are constants and
$E\eta_{\mu}\eta_{\nu}=\delta_{\mu\nu}$. We denote
$$
<q_1q_2>=A,~<p_1q_2>=B,~<q_1p_2>=C,~<p_1p_2>=D.
$$
A solution of Eqs (\ref{eq:EPR7}) may be given for example by
$$
f_1=A\eta_1,~~f_2=\eta_1,
$$
$$
g_1=B\eta_1+(D-\frac{BC}{A})\eta_2,~g_2=\frac{C}{A}\eta_1+\eta_2
$$
Hence the representation of the quantum correlation function in
terms of the separated classical random processes  (\ref{eq:EPR5})
is proved.

The condition of reality of the functions $ \xi_n (\alpha_n,\lambda)$ is important.
It means that the range of $ \xi_n (\alpha_n,\lambda) $ is the set of eigenvalues
of the operator $q_n(\alpha_n).$ If we relax this condition then one can get a
hidden variable representation just by using an expansion of unity:
$$
\left<\psi|q_1 (\alpha_1)
q_2(\alpha_2)|\psi\right>=\sum_{\lambda}\left<\psi|q_1 (\alpha_1)|\lambda\right>
\left<\lambda|q_2(\alpha_2)|\psi\right>
$$
For a discussion of this point in the context of a noncommutative
spectral theory see \cite{Vol2}.

Similarly one can prove a representation
\begin{equation}
\label{eq:EPR8} \left<\psi|q_1 (t_1) \otimes
q_2(t_2)|\psi\right>=\int \xi_1 (t_1,\lambda) \xi_2(t_2,\lambda)
d\rho(\lambda)
\end{equation}
where $q_n(t)=q_n+p_nt,~n=1,2$ is a free quantum evolution of the
particles. It is enough to take
$$
\xi_1(t_1,\lambda)=f_1(\lambda)+g_1(\lambda)t_1,
~~\xi_2(t_2,\lambda)=f_2(\lambda)+g_2(\lambda)t_2.
$$

To summarize, it is shown in the note that, in contrast to the
Bell`s theorem for  the spin or polarization variables, for the
original EPR correlation functions which deal with positions and
momenta one can get a local realistic representation in terms of
separated random processes. The representation is obtained for any
state including entangled states. Therefore the original EPR model
does not lead to quantum nonlocality in the sense of Bell even for
entangled states. One can get quantum nonlocality in the EPR
situation only if we (rather artificially) restrict ourself in the
measurements with a two dimensional subspace of the infinite
dimensional Hilbert space corresponding to the position or
momentum observables.

If we adopt the Bell approach to the local realism (i.e. through
classical  separated stochastic processes) then
one can say that the original EPR model admits a local realistic
description in contrast to what was expected for the model. It
follows also that the phenomena of quantum nonlocality in the
sense of Bell depends not only on the properties  of entangled
states but also on  particular observables which we want to measure
(bounded spin-like or unbounded momentum and position
observables). An interrelation of the roles of entangled states
and the bounded observables in  considerations of local realism
and quantum nonlocality deserves a further theoretical and experimental study.

Discussion of appropriate experiments  is  out the scope of the present paper.
Here our primary goal was to show that the original EPR model admits a local
realisitic representation in the sense of Bell for any states 
if we measure positions and momenta.
In particular it  would be interesting to explore the transition from EPR locality
to quantum nonlocality when we measure only spin-like observables by using
states produced in a pulsed nondegenerate optical parametric amplifier (NOPA).
To this end one has to study correlations of operators which interpolate
between the bounded spin-like or parity-like observables and position and momenta operators.  

{ \bf Acknowledgments}

The paper was supported by the grant of The Swedish Royal Academy of Sciences
on collaboration with scientists of former Soviet Union, EU-network on quantum probability and applications
and RFFI-0201-01084.

\end{document}